\documentclass[preprint showpacs,preprintnumbers,amsmath,amssymb]{revtex4}
\usepackage{xcolor}
\usepackage{graphicx}% Include figure files
\usepackage{dcolumn}% Align table columns on decimal point
\usepackage{bm}% bold math
\usepackage{multirow}
\usepackage{natbib}
\usepackage{color}
\usepackage{cases} 

\newcommand{\Vmat}{{\boldmath $\cal V$}}
\newcommand{\Kmat}{{\boldmath $\cal K$}}

\newcommand{\numat}{\mbox{\boldmath $\nu$}}

\newcommand{\Smat}{\mbox{\boldmath $\cal S$}}
\newcommand{\SSmat}{\mbox{\boldmath $S$}}
\newcommand{\Xmat}{\mbox{\boldmath $X$}}

\begin{document}
\preprint{APS/123-QED}

%\title{ Optical shielding of ultracold K-Cs binary collision}
\title{Low-energy electron impact dissociative recombination and vibrational transitions of N$_2^+$}

\author{A. Abdoulanziz$^{1}$}
\author{C. Argentin$^{1,2}$}
\author{V. Laporta$^{3}$}
\author{K. Chakrabarti$^{4}$}
\author{A. Bultel$^{2}$}
\author{J. Tennyson$^{5,1}$}
\author{I. F. Schneider$^{1,6}$}
\author{J. Zs Mezei$^{7,1}$}
\email[]{mezei.zsolt@atomki.hu}
\affiliation{$^{1}$LOMC CNRS-UMR6294, Universit\'e le Havre Normandie, F-76058 Le Havre, France}%
\affiliation{$^{2}$CORIA-UMR6614 CNRS-Universit\'e de Rouen Normandie, 76800 Saint-Etienne du Rouvray, France}%
\affiliation{$^{3}$Istituto per la Scienza e Tecnologia dei Plasmi, CNR, 70126 Bari, Italy}%
\affiliation{$^{4}$Dept. of Mathematics, Scottish Church College, 700006 Kolkata, India}%
\affiliation{$^{5}$Dept. of Physics and Astronomy, University College London, WC1E 6BT London, UK}%
\affiliation{$^{6}$LAC CNRS-FRE2038, Universit\'e Paris-Saclay, F-91405 Orsay, France}%
\affiliation{$^{7}$Institute for Nuclear Research (ATOMKI), H-4001 Debrecen, Hungary}%
\date{\today}

\begin{abstract}
Cross sections and thermal rate coefficients are computed for electron-impact dissociative recombination and vibrational excitation/de-excitation of the N$_2^+$ molecular ion in its lowest six vibrational levels, for collision energies/temperatures up to 2.3 eV/5000 K. 
\end{abstract}

%\pacs{33.80. -b, 42.50. Hz}% PACS, the Physics and Astronomy
                             % Classification Scheme.
%\keywords{coupled-channel, optical shielding, KCs}%Use showkeys class option if keyword

                              %display desired
\maketitle

\section{\label{sec:intro}Introduction}

The nitrogen molecule N$_2$ is one of the most widely studied species so far in plasma physics. 
Being very stable at low temperature, it is very abundant in the Earth atmosphere, 
and is notably present in other planetary atmospheres 
- Titan 98.4~\%~\cite{Krasnopolsky_2014}, 
Triton~\cite{Elliot_2000} 
Pluto~\cite{Krasnopolsky_2020},
Venus 3.5~\% and Mars 1.9~\%~\cite{Krasnopolsky_2014}. 
For other trans-Neptunian objects than Pluto, this molecule is also one of the main component of the 
ices - spectroscopically observed at their surfaces - and may produce a very thin atmosphere when the
temperature increases under solar irradiation~\cite{Young_2020}. Under the influence of an electric field, 
high altitude planetary atmosphere can be crossed by  giant discharges of a few milliseconds duration called sprites, 
whose spectroscopic signature is mainly due to spontaneous emission from N$_2$ excited  electronic states~\cite{Armstrong_2000}. The application of N$_2$ as seeding gas in magnetically confined fusion plasmas (ITER and JET equipments) will help in the reduction of power loads on the tungsten divertor region. Nitrogen may be preferable as an extrinsic radiator to noble gases (neon) as it mostly affects the divertor radiation without significantly increasing core radiation~\cite{Oberkofler2013,Giroud2010}. 

All this facts justify the various  studies of the role 
of the N$_2$ molecule in the cold plasmas, from the state-to-state description of its electron impact induced reactivity~\cite{ Laporta2012, Laporta2014, Kitajima2017} to the detailed modeling of its contribution to the plasma kinetics~\cite{Guerra2004, Panesi2011, Capitelli2014, Heritier2014}.

Consequently, the N$_2^+$ cation is also of huge interest. 
Due to the solar irradiation, the production of N$_2^+$ on excited vibrational states plays a significant role in the characteristics of the Earth's thermosphere~\cite{Torr_1983}. 
It is also the main molecular cation in the atmosphere of Titan~\cite{Lammer2000} and Triton~\cite{Yung1990}. 
On the other hand, during the atmospheric entry of a spacecraft in Earth's and Titan's atmospheres, the hypersonic compression of the gases leads to the formation of a plasma departing from local thermodynamic equilibrium~\cite{Annaloro_2014}. The ionic composition, including N$_2^+$, plays a key role in the radiation emitted by the plasma in the near UV spectral region~\cite{Plastinin_2007}. 
In many 
plasma-assisted industrial processes elaborated so far, 
the plasma reactivity is greatly enhanced by the presence of  N$_2^+$. This is, for instance, the case in the ammonia synthesis in plasmas/liquid processes~\cite{Sakakura_2019}. N$_2^+$ is also very effective in the antibacterial treatment of polyurethane surfaces~\cite{Morozov_2016}. 
Moreover, N$_2^+$ -- like N$_2$ -- is a key ingredient in the steel nitriding, resulting in the improving of its frictional wear resistance, surface hardness and corrosion resistance~\cite{Sharma_2008}. Furthermore, N$_2^+$ also plays a major role in the dermatological treatments based on the nitrogen radio-frequency discharges~\cite{Holcomb_2020}.

The characteristics of the nitrogen-containing plasmas cannot be fully understood without a deep knowledge of the 
reactivity of N$_2^+$, in particular by collisions with electrons.

Dissociative Recombination (DR) is the major molecular cation destruction reaction, that takes place when an electron collides with the N$_2^+$ 
molecular cation, leading to neutral atomic fragments: 
\begin{equation}
\mbox{N}_2^{+}(v_{i}^{+}) + e^-(\varepsilon) \longrightarrow \mbox{N} + \mbox{N}. 
\label{eq:DR}
\end{equation}
Here $\varepsilon$ is kinetic energy of the incident electron and $v_{i}^{+}$ the initial vibrational quantum number of the target.
Alongside DR, other competitive processes can occur: 
\begin{equation}
\mbox{N}_2^+(v_{i}^{+}) + e^{-}(\varepsilon) \longrightarrow  \mbox{N}_2^{+} (v_{f}^{+}) + e^{-}(\varepsilon _{f}), \label{eq:VE}
\end{equation}
i.e. elastic (EC) $(v_f^+ =v_i^+ )$, inelastic (IC)  $(v_f^+ >v_i^+ )$ and super-elastic (SEC) collisions $(v_f^+ < v_i^+ )$, $v_{f}^{+}$ standing for the 
final vibrational quantum number of the target ion. These processes are also known as Elastic Scattering (ES), Vibrational Excitation (VE) and Vibrational deExcitation (VdE) respectively.

The elementary non-thermal electron driven processes, in particular dissociative recombination, has been studied experimentally  using plasmas with laser induced photo-fluorescence techniques~\cite{Zipf1980},  shock tubes~\cite{Cunningham1972}, discharge afterglow experiments~\cite{mitra1953,Kaplan1948} and microwave techniques~\cite{Biondi1949}. The most detailed collisional data can be obtained in merged beam~~\cite{Noren1989} and/or storage ring experiments~\cite{Peterson1998}.

Two different sets of theoretical calculations have been performed~\cite{Little2014,Guberman2012,Guberman2013,Guberman2014} 
on the DR of N$_2^+$. They involved different underlying quantum chemistry but rather similar nuclear dynamics calculations;
both these studies  focused on the ground and the lowest three vibrational levels of the N$_2^+$ target.

%\red{
While both results show good agreement with experiment for the ground vibrational level, the rates for the higher vibrational levels calculated in~\cite{Little2014}, contrary to those of~\cite{Guberman2012,Guberman2013,Guberman2014}, indicate less strong vibrational dependence on temperature, in agreement with the experimental results.
%} 

Our aim with this paper is to extend as far as possible the calculations started in ~\cite{Little2014}. 
This extension refers to:

i) the kinetic energy of the incoming electron: up to $2.3$ eV {\it vs} 1 eV previously.

ii) the elementary processes explored: besides the DR studied in the past, the VE and VdE cross sections and rate coefficients are computed.

iii) the vibrational levels considered in the vibrational transitions: up to the fifth excited level of the target  {\it vs} the third previously, and the lowest ten vibrational levels as final ones.

\noindent The rotational effects have been neglected, since they are important only at very low collision energies.

All these extensions make our results relevant for the atmospheric and cold plasma environments, at electron temperatures where the rotational effects can be neglected.

The paper is organized as follows: After a brief description of the theoretical approach (section~\ref{sec:theory}), we present in more details the molecular data used in the calculations (section~\ref{sec:moldata}) followed by the presentation of the results (section~\ref{sec:res}). The paper is ended by conclusions. 

\section{Theoretical approach}\label{sec:theory}

The efficiency of our method of modeling the electron/diatomic cation collisions, based on the Multichannel Quantum Defect Theory (MQDT) has been proved in many previous studies on different species, including H$_2^+$ and its isotopologues~\cite{Chakrabarti2013,Motapon2014,Epee2016}, ArH$^+$~\cite{Abdoulanziz2018}, CH$^+$~\cite{Mezei2019}, SH$^+$~\cite{Kashinski2017},  etc.  
The general ideas of our approach were already presented in detail in our previous study of the N$_2$$^+$ dissociative recombination ~\cite{Little2014} and, therefore, here we restrict ourselves to its major steps.

The reactive processes~(\ref{eq:DR}) and (\ref{eq:VE}) involve \textit{ionization} channels - describing the scattering of an electron on the target cation - and \textit{dissociation} channels - accounting for atom-atom scattering. The mixing of these channels results in quantum interference of the \textit{direct} mechanism - in which the capture takes place into a doubly excited dissociative state of the neutral system - and the \textit{indirect} one - in which the capture occurs via a Rydberg bound state of the molecule belonging to a \textit{closed} channel, this state being predissociated by the dissociative one. 
In both mechanisms the autoionization - based on the existence of \textit{open}  ionization channels - is in competition with the predissociation, and can lead to the excitation or to the de-excitation of the cation.

One starts with the building of the {\it interaction matrix} {\Vmat} that drives the collision, whose elements quantify the couplings between  the different channels - ionization and/or dissociation ones. 

\begin{figure}[t]
	\begin{center}
		\centering
		\includegraphics[width=0.75\linewidth]{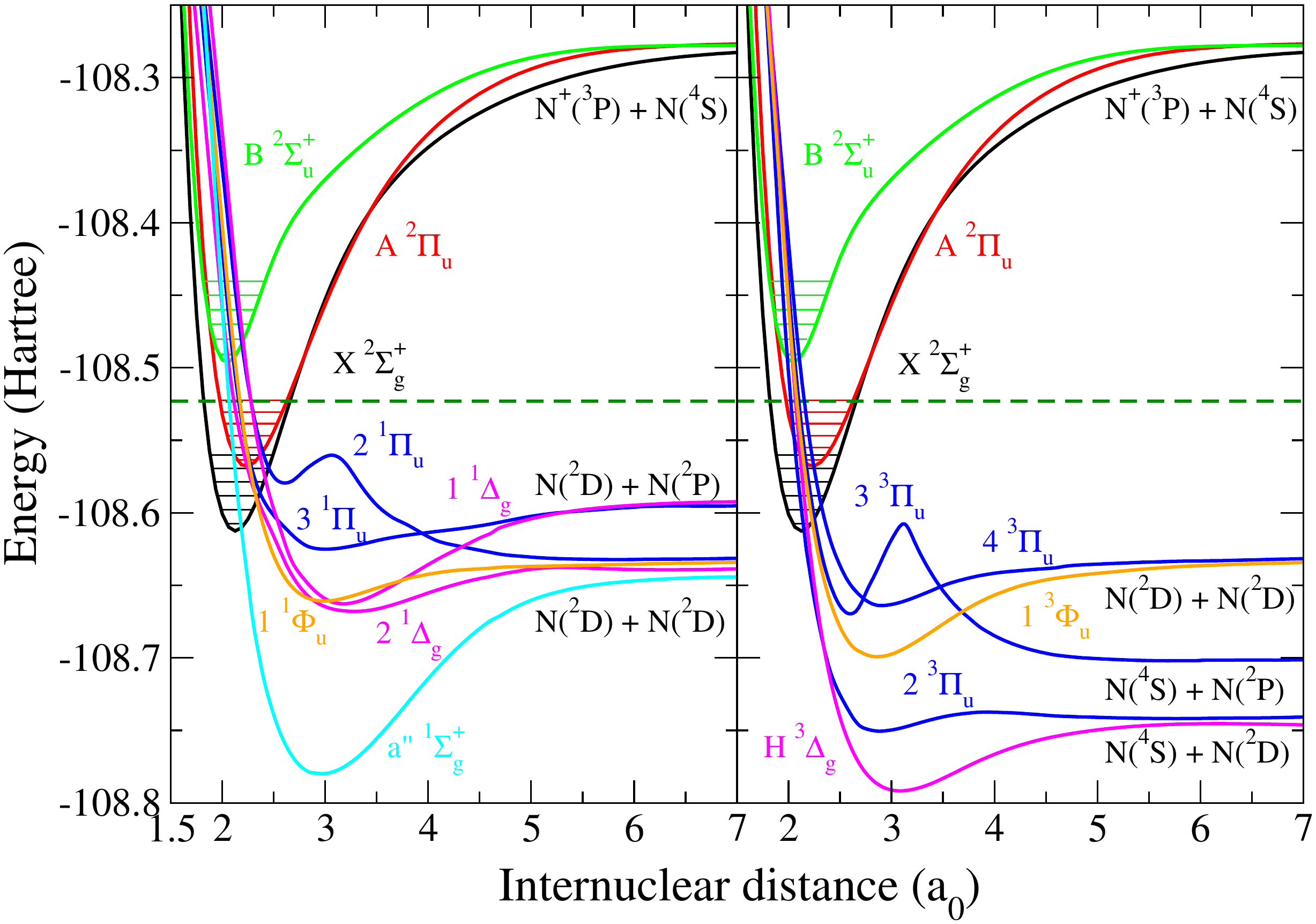}
		\caption{Potential energy curves (PEC) relevant for the DR of N$_2^+$~\cite{Little2014}. 
Target cation N$_2^+$: ground electronic state ($X^2\Sigma_g^{+}$) black,
first excited state ($ A^2\Pi_u$) red, 
second excited state $B ^2\Sigma_u^{+}$ green. 
Neutral system N$_2$: 
Left panel, singlet states of different symmetries - blue for $^1\Pi_u$, magenta for $^1\Delta_g$, cyan for $a"^1\Sigma_g^{+} $ and orange for $^1\Phi_u$. 
Right panel, triplet states of different symmetries - blue for $^3\Pi_u$, magenta for $H^3\Delta_g$ and orange for $^3\Phi_u$. 
The lowest five vibrational levels of each electronic state of the ion and the dissociative asymptotic limits for all states are shown. The green dashed line gives the upper limit of the total energy of the system, above which our results are still reasonably correct (see text).}
		\label{fig:1}
	\end{center}
\end{figure}

More specifically, each of the ionization channels, built on the N$_2^+$ ion in one of its three lowest electronic states  
- $X$ ${^2}\Sigma{_g^+}$, $A$ ${^2}\Pi{_u}$ or $B$ ${^2}\Sigma{_u^+}$, see Fig. 1 - in a particular vibrational level, interacts not only with all the dissociation exit channels (Rydberg-valence interaction), but also with the other ionization channels (Rydberg-Rydberg interactions) - Fig. 2. 
Depending on the total energy of the system 
these ionization channels can be {\it open} - either as entrance channels, describing the incident electron colliding the ion in its ground electronic state, or  exit channels, describing the auto-ionization, i.e. elastic scattering, vibrational excitation and de-excitation  - or {\it closed} - describing the resonant temporary captures into Rydberg states. 

Once the {\Vmat}-matrix is elaborated, we build the short-range reaction matrix {\Kmat} of the collision, as a second order perturbative solution of the Lippmann-Schwinger equation. The diagonalized version of the {\Kmat}-matrix (in the eigenchannel representation) whose eigenvalues are expressed in terms of long range phase-shifts of the eigenfunctions, together with the vibronic couplings between the ionization channels, serve for the building of the frame transformation matrices. 

Applying a Cayley transformation on these latter matrices we can set up the generalized scattering matrix {\Xmat}.
 The Seaton's method of 'eliminating' the closed channels ~\cite{Seaton1983} is then employed, resulting in the physical scattering matrix {\Smat}:

\begin{equation}
\SSmat = \Xmat_{oo}-\Xmat_{oc}\frac{1}{\Xmat_{cc}-\exp({\rm -i 2 \pi} \numat)} \Xmat_{co}\,,
\label{eq:elimination}
\end{equation}
relying on the block-matrices involving open ({\Xmat$_{oo}$}), open and closed ({\Xmat$_{oc}$} and {\Xmat$_{co}$}) and closed (\Xmat$_{cc}$) channels. The diagonal matrix {\numat} in the denominator of equation (\ref{eq:elimination}) contains the effective quantum numbers corresponding  to the  the vibrational thresholds of the closed ionisation channels at given total energy of the system.

Finally, the cross section for the dissociative recombination and for the vibrational transitions - vibrational excitation/ 
de-excitation
and elastic scattering 
write respectively as:

\begin{equation}
\sigma_{diss \leftarrow v_{i}^{+}} 
= \frac{\pi}{4\varepsilon} \rho^{sym} \sum_{l,j } \left| S^{\Lambda}_{d_j,lv_{i}^{+}}\right|^2 \nonumber \\
\label{eq:csDR}
\end{equation}
and 
\begin{equation}
\sigma_{v_f^+ \leftarrow v_i^{+}} = \frac{\pi}{4\varepsilon} \rho^{sym}
\sum_{l,l'}\left| S^{\Lambda}_{l'v_f^+,lv_i^+} - \delta_{l,l'}\delta_{v_i^+,v_f^+} \right |^2\,. \label{eq:csVE-partial}
\end{equation}
where $d_j$ stands for a given dissociative state and $\rho^{sym}$ the ratio between the state-multiplicities of the neutral and the target ion.  

\begin{figure}[t]
	\begin{center}
		\centering
		\includegraphics[width=.75\linewidth]{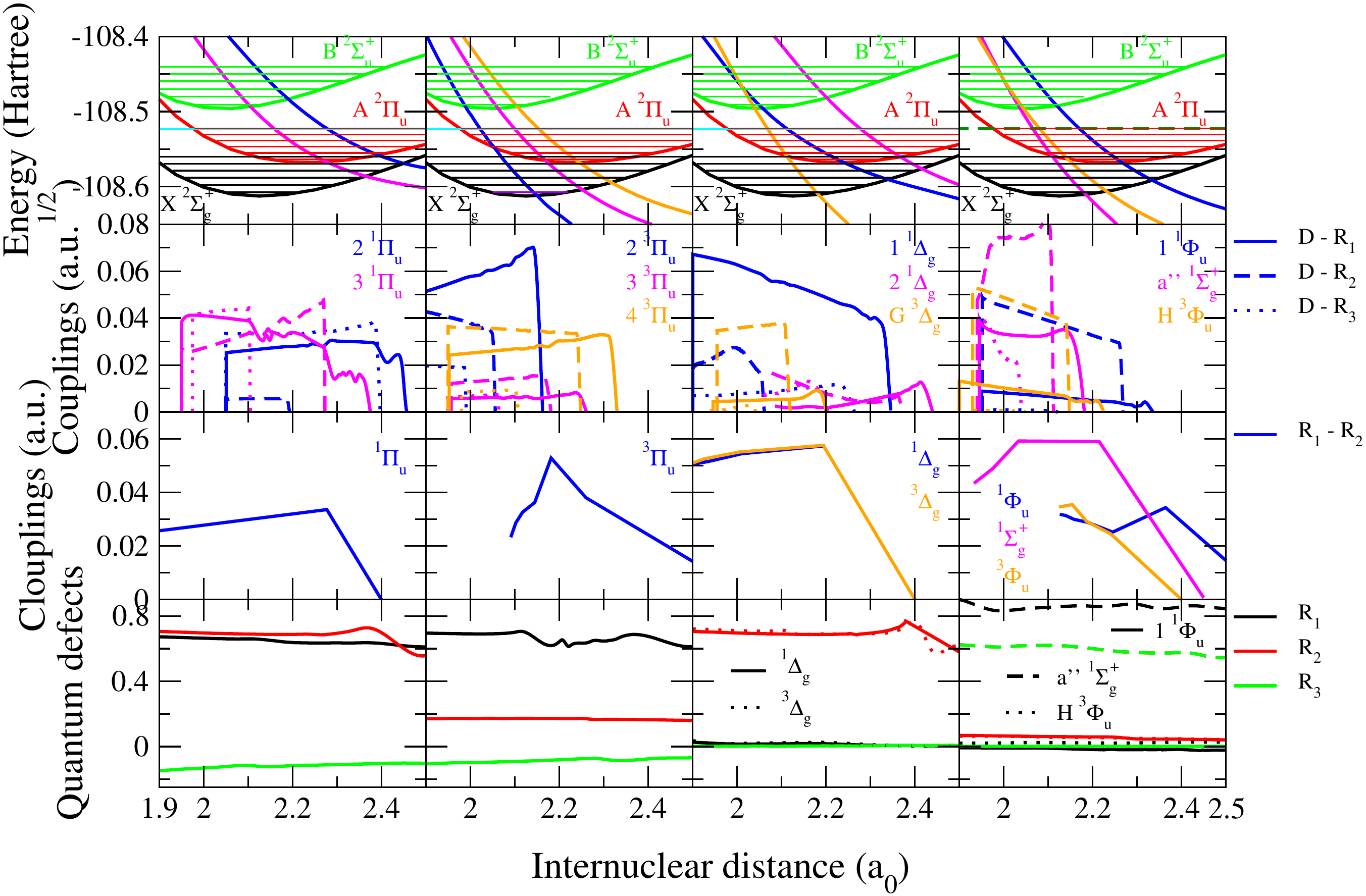}
		\caption{Molecular data sets for the modeling of reactive collisions between electrons and N$_2^+$~\cite{Little2014}.
$1^{st}$ row: PECs of the relevant states of the ion and of the neutral for all relevant symmetries. $2^{nd}$ row: Rydberg-valence electronic couplings. $3^{rd}$ row: Rydberg-Rydberg electronic couplings. $4^{th}$ row: Quantum defects characterizing the Rydberg series of states. }
		\label{fig:2}
	\end{center}
\end{figure}

\section{Molecular data \label{sec:moldata}}

The nuclear dynamics in low-energy electron/molecular cation collisions crucially depends on the molecular structure of the 
target and of the formed neutral - often superexcited - complex. 
The 
relevant molecular data sets 
consist in 
the potential energy curves (PECs) of the target cation - for the ground and for the excited electronic states - 
the PECs of the 
doubly excited bound or dissociative molecular states of the neutral, 
the quantum defect-functions characterizing the bound mono-excited Rydbers series of the neutral, 
and 
the coupling functions between the several - ionization and dissociation - continua.

One of the few quantum chemistry methods capable of producing the highly-excited 
molecular states at the required accuracy is based on the R-Matrix 
Theory~\cite{Tennyson_PR_2010}. Bound and resonant adiabatic potential 
energy curves of the 
valence and Rydberg states of N$_2$ having singlet and triplet symmetries 
were obtained by Little and Tennyson\cite{Little_bound_state,Little-resonance_state} using R-matrix
calculations on fine grid of internuclear separations. The 
diabatic curves, couplings and quantum defects relevant for the dissociative 
recombination of N$_2^+$ were presented in~\cite{Little2014}. The electronic 
states of the target were calculated using the standard quantum chemistry 
program suite Molpro~\cite{MOLPRO-WIRE}.
Figure~\ref{fig:1} shows the PECs of the dissociative molecular states of N$_2$, 
as well as those of the relevant states of N$_2^+$,
involved in our previous~\cite{Little2014} 
and present calculations. 

The same PECs are displayed by symmetries on the first row of Figure~\ref{fig:2},  
which contains the whole ensemble of molecular data relevant for the modeling of the internuclear dynamics. 
Whereas its first row illustrates how favorable the crossings are between the PECs of the dissociative states with those of the target ones - i.e. the Franck-Condon effect -
the driving interactions of the dynamics - the Rydberg-valence couplings - are shown in the second row. The third row gives the Rydberg-Rydberg couplings: 
In the present calculation, only the couplings among the series correlating to the ground ($X$) and first excited ($A$) state of the ion have been considered.                                                                                                                                            
And finally the last row of the figure displays the quantum defects characterizing the Rydberg series built, each of them, on one of the three cores X, A and B.

\section{Results and discussions \label{sec:res}}

Based on the molecular data already presented in fig.~\ref{fig:2},
we have performed the nuclear dynamics calculations using the MQDT approach presented in Section~\ref{sec:theory}.
The DR, VE and VdE cross sections have been calculated considering the N$_2^+$ target in one of its lowest six vibrational states, and focusing on the vibrational transitions to the lowest ten vibrational levels, when energetically accessible.
Table~\ref{table:1} shows the energies of these latter levels 
relative to $v_i^+ = 0$ of the target. 

\begin{table}[t]
\caption{The energies of the vibrational levels of the N$_2^{+}$ molecular cation - relative to the ground one - involved either as initial or as final levels in the present calculations. }
\label{table:1}
\footnotesize
\begin{center}
\begin{tabular}{c|cccccccccc}
	\hline\hline
$v^+$ & 0 & 1 & 2 & 3 & 4 & 5 & 6 & 7 & 8 & 9 \\[2pt]
\hline
$E_{v^+}$ (eV) & 0.0 & 0.266 & 0.528 & 0.786 & 1.040 & 1.290 & 1.536 & 1.777 & 2.014 & 2.248\\[3pt]
\hline\hline
\end{tabular}
\end{center}
\end{table}
	
The calculations have been performed by taking into account both the {\it direct} 
and the {\it indirect} 
mechanisms, the reaction matrix being evaluated in the second order,
and all their vibrational levels - $81$, $66$ and $50$ respectively, associated to  {\it open} or  {\it closed} ionization channels, according to the total energy of the system - have  been fully accounted. 

The cross sections have been calculated for all the relevant symmetries listed in figs.~\ref{fig:1} and \ref{fig:2}, for collision energies of the incident electron ranging between $10^{-5}$ and $2.3$ eV, with an energy step of $0.01$ meV. These cross sections have been summed up to obtain the global cross sections.

%\begin{landscape}
 \begin{figure*}[t]
	\centering
	\includegraphics[width=0.75\linewidth]{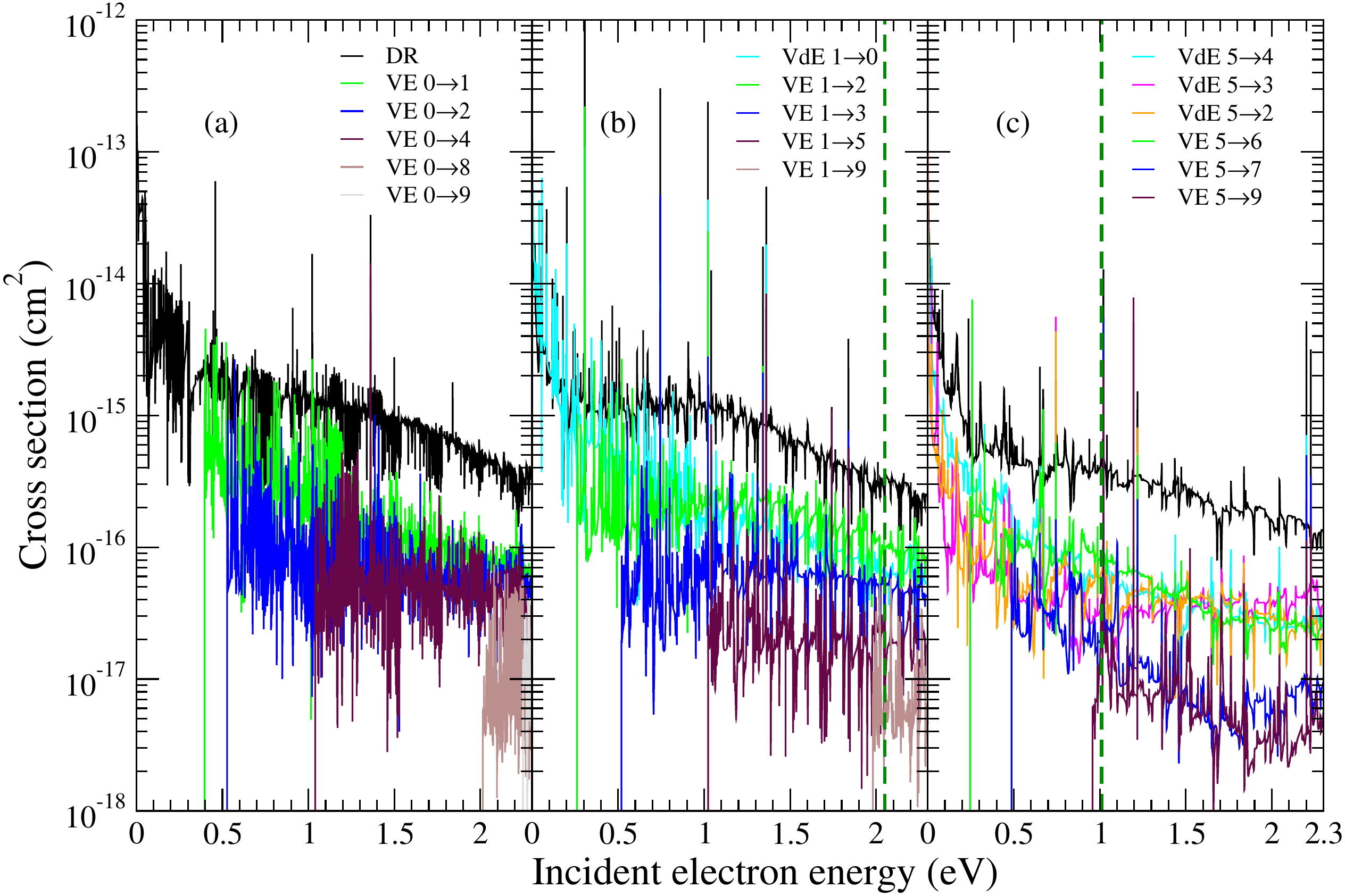}
	%\centering
	\caption{Global DR, VE and VdE cross sections of the N$_2^+$ $v_i^+ = 0$ (a), $v_i^+ = 1$ (b) and $v_i^+ = 5$ (c) %panels 
	as a function of the collision energy. For vibrational transitions (VE and VdE) we label the processes as transitions from the initial to the final vibrational levels of the target.  
The vertical dashed dark-green line gives the precision limit of the calculations (for details see text).}
	\label{fig:3}
\end{figure*}
%\end{landscape}

The global DR, VE and VdE cross sections for target cations having initial vibrational levels $v_i^+=0,1$ and $5$ are shown in figure \ref{fig:3} (a), (b) and (c) panels respectively. The vertical dark-green dashed lines in the mid and upper panels of figure~\ref{fig:3} mark the energy below which the calculations are the most accurate. Above these thresholds the calculations neglect the role of the higher lying dissociative states of the neutral.   

Nevertheless, the data displayed continue to be reasonably correct above these thresholds because these dissociative states penetrate into the ionization continuum well above these thresholds, forming favourable/non-vanishing Franck-Condon overlaps with the target electronic states 
at even higher collision energies. This Franck-Condon overlap is proportional with the first order term of the direct cross section.  In addition the couplings of these dissociative states with the Rydberg series are generally weaker, leading to less important cross sections in second order.

The {\it direct} mechanism is responsible for the background  $1/E$ behaviour of the cross sections, while the {\it indirect} one through the temporary capture into the Rydberg states produces all the resonance structures dominating the cross sections.

Among all the processes studied here dissociative recombination (black curves in fig.~\ref{fig:3}) predominates. %, their cross sections being higher (with a factor of 2-3 for higher collision energies) than those obtained for the vibrational transitions (coloured curves). 
The global DR cross section increases as we change the initial vibrational state of the target by unity and starts to decreases as we arrive at $v_i^+=5$.  While the vibrational de-excitation (cyan curves for initial vibrational levels higher than $0$) are in competition with the  DR cross section, at higher collision energies their overall cross section values are at least with a 
factor of $5$ smaller than those of the DR. The vibrational excitations (green, blue, violet, maroon, etc. curves) show threshold effects at the collision energies where they become open. Moreover, one can see that for a given initial vibrational level $v_i^+$ the $|\Delta v^+|=1$ vibrational transitions are the most probable ones, decreasing monotonically with $|\Delta v^+|$
for the transitions between more distant levels.

\begin{figure*}[t]
	\centering
	\includegraphics[width=0.75\linewidth]{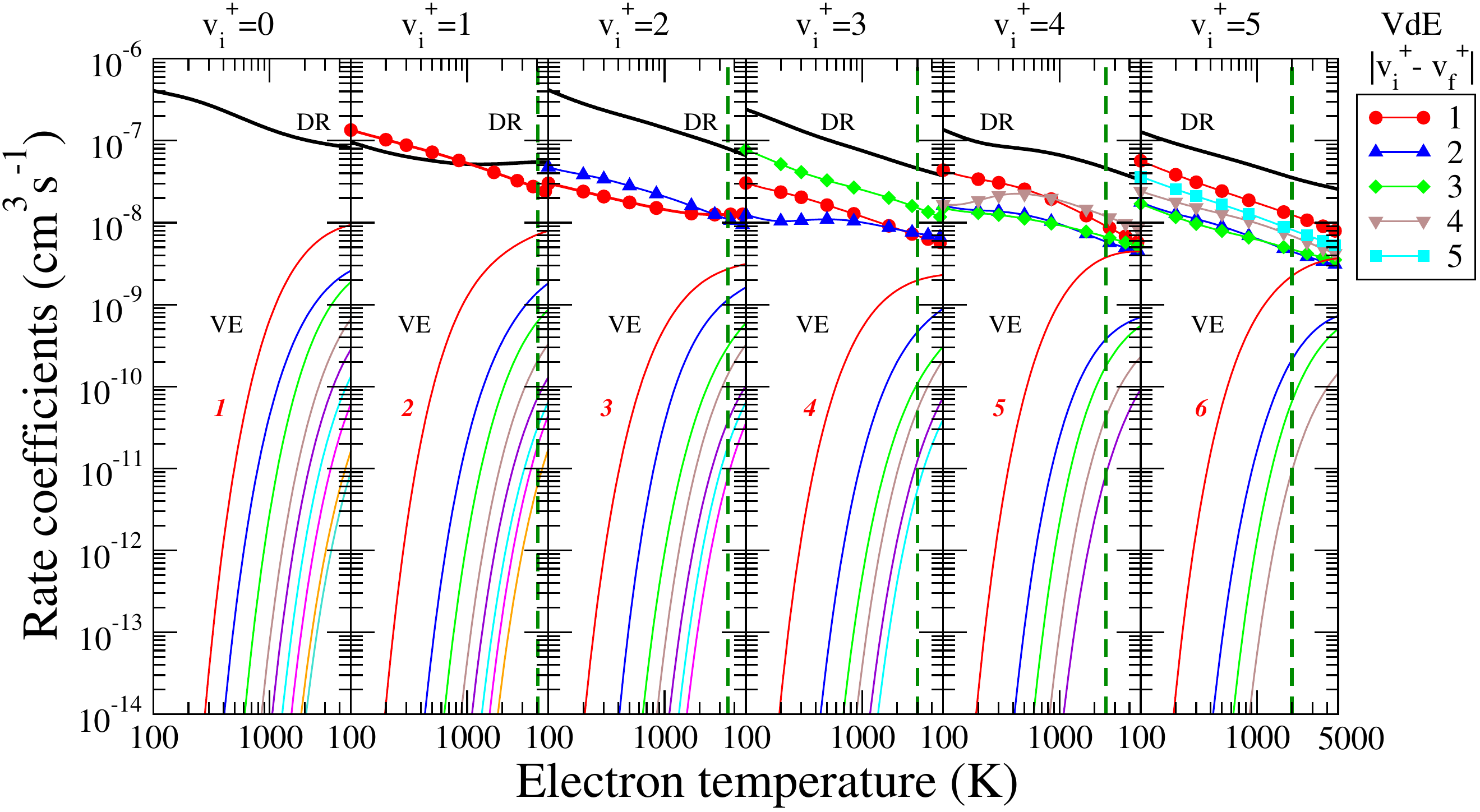}
	\caption{
	Maxwell rate coefficients for all the relevant electron-induced processes on  N$_2^+$ initially on $v_i^+ = 0-5$ vibrational levels :  Dissociative recombination (black line), vibrational excitation (thin coloured lines) and vibrational de-excitation (symbols and thick coloured lines). For the vibrational excitations all the transitions are shown up to $v^+_f=9$  
	with the lowest transition being labeled on each figure. The excitation and the de-excitation up to the final vibrational quantum numbers are given. The green dashed line gives the precision limit of our calculation given in temperatures (for details see text).
	}
	\label{fig:6}
\end{figure*}

Figure~\ref{fig:6} shows the thermal rate coefficients of all processes for the six lowest initial vibrational levels of N$_2^+$. The green dashed line gives the precision limits of our calculation expressed now in electron temperatures.

The DR (solid black line in figure~\ref{fig:6}) and VdE (symbols and thick coloured lines) rate coefficients 
decrease monotonically 
with the temperature, while the VE (thin coloured lines) ones 
increase,
partly because of the threshold behaviour of their corresponding cross sections.
The largest rate coefficients we obtained are those for the DR. With the exception of the $v_i^+=1$ case the VdE rate coefficients are smaller than those for the DR. At $v_i^+=1$, the DR is in competition with VdE but, for $v_i^+>1$ DR exceeds VdE with a factor of $2-5$. We can see from figure~\ref{fig:6} that the VE process is relatively important at high electron temperatures only. Moreover, higher we go with the initial vibrational quantum number of the target cation, more probable VE becomes.

\begin{table}[t]
	\renewcommand{\arraystretch}{1.00}
	\centering
	\caption{\label{tab:2}List of the fitting parameters used in formula~(\ref{eqn:fitt}), temperature regions and root mean squares
for the DR rate coefficients of N$_2^+$ ($v_i^+=0-5$).}
	\footnotesize
%	\scriptsize
	\begin{tabular}{ccccccc}
		\hline
	&	$v_i^+$ &  Temperature range & $A$                      & $\alpha$                  &  $B$ & RMS \\
	&	 &  K & $(cm^3s^{-1}K^{-\alpha})$                      &                   &  $(K)$ & \\
		\hline
%EC	&	0    & 100$\le$T$\le$5000 & 3.37355$\times10^{-5}$ & -0.498093 & -1.97069 & 0.0095 \\
%	&	1     & 100$<$T$\le$5000 & 3.35449$\times10^{-5}$ & -0.487562 & -2.85638 & 0.0086 \\
%	&	2     & 100$<$T$\le$5000 & 4.29406$\times10^{-5}$ & -0.516353 & 1.01360 & 0.0050 \\
%	&	3     & 100$<$T$\le$5000 & 4.32596$\times10^{-5}$ & -0.512870 & 3.42679 & 0.0023 \\
%	&	4     & 100$<$T$\le$5000 & 4.53781$\times10^{-5}$ & -0.519476 & 1.46782 & 0.0029 \\
%	&	5     & 100$<$T$\le$5000 & 4.32166$\times10^{-5}$ & -0.511299 & 2.01044 & 0.0022 \\
%		\hline
DR	&	0    & 100$\le$T$\le$700 & 1.56020$\times10^{-5}$ & -0.679449 & 53.0333 & 0.0051 \\
	&	      & 700$<$T$\le$5000 & 2.27099$\times10^{-7}$ & -0.129049 & -388.399 & 0.0040 \\
	&	1    & 100$\le$T$\le$900 & 1.11178$\times10^{-7}$ & -0.120373 & -41.6843 & 0.0049 \\
	&	      & 900$<$T$\le$5000 & 2.60704$\times10^{-8}$ & 0.086645 & -85.4307 & 0.0055 \\
	&	2    & 100$\le$T$\le$1000 & 1.54430$\times10^{-6}$ & -0.345836 & -28.1712 & 0.0053 \\
	&	      & 1000$<$T$\le$5000 & 7.09592$\times10^{-6}$ & -0.544674 & 145.630 & 0.0030 \\
	&	3    & 100$\le$T$\le$5000 & 1.53033$\times10^{-6}$ & -0.438705 & -19.1368 & 0.0088 \\
	&	4    & 100$\le$T$\le$460 & 5.25573$\times10^{-8}$ & 0.042384 & -76.5345 & 0.0032 \\
	&	      & 460$<$T$\le$5000 & 3.42600$\times10^{-6}$ & -0.539469 & 208.028 & 0.0031 \\
	&	5    & 100$\le$T$\le$5000 & 7.20627$\times10^{-7}$ & -0.394354 & -7.90548 & 0.0106 \\
		\hline
	\end{tabular}
	\normalsize
\end{table}

And finally, in order to allow the versatile implementation of the rate coefficients  in kinetics modelling codes, we have fitted them with 
Arrhenius-type formulas. The calculated rate coefficients for the dissociative recombination 
%and for the elastic collisions 
of electrons with N$_2^+$ on each of its lowest $6$ vibrational levels ($v_i^+=0, 1, ..., 5$) have been interpolated under the mathematical form:

\begin{equation}\label{eqn:fitt}
k^{fitt}(T) = A \, T^{\alpha} \, \exp\left[-\frac{B}{T}\right]
%\label{kec}
\end{equation}

%\begin{equation}\label{eqn:N2_EC_Interpolation}
%k_{v_i^+}^{EC,DR}(T_e) = A_{v_i^+} \, T_e^{\alpha_{v_i^+}} \, \exp\left[-\frac{B_{v_i^+}}{T_e}\right]
%\label{kec}
%\end{equation}
%
%
%	\begin{figure}[b]
%	\centering
%	\includegraphics[width=0.88\linewidth]{reldiff}
%	\caption{Relative difference between  the fitted - with formulas (\ref{eqn:N2_EC_Interpolation}), (\ref{eqn:N2_VE_VdE_Interpolation}) - and calculated Maxwell rate coefficients for $v_i^+=1$. 
%	}
%	\label{fig:7}
%\end{figure}
%
\noindent over the electron temperature range 100~K~$\le T_e\le$~5~000~K and/or for rate coefficients larger then $10^{-14}$ cm$^3$s$^{-1}$, as displayed in fig.~\ref{fig:6}.
The $A$, $\alpha$ and $B$ fitting parameters used in equation~(\ref{eqn:fitt}) together with the temperature regions are listed in table~\ref{tab:2} for DR, and in table~\ref{tab:3} for VE and VdE processes. The efficiency of the fitting are characterised with the Root Mean Squares, and we were able to reproduce the MQDT rate coefficients with a precision higher than $97\%$.
%inacuracy of less than 3$\%$. 
%pression less the 3$\%$. 

\begin{table*}[t]
	\renewcommand{\arraystretch}{1.00}
	\centering
	\caption{\label{tab:3}List of the fitting parameters used in formula~(\ref{eqn:fitt}), temperature regions and root mean squares for the VE and VdE rate coefficients of N$_2^+$  ($v_i^+=0-5$ and $v_f^+=9$). The lines having bold $v_f^+$ values belong to VdE.}
%	\footnotesize
	\scriptsize
	\begin{tabular}{ccccccc}
		\hline
	$v_i^+$ &	$v_f^+$ & Temperature range & $A$             & $\alpha$         & $B$ & RMS \\
	 &	 &   (K) & $(cm^3s^{-1}K^{-\alpha})$   &      & $(K)$ & \\
		\hline
0 	&	1    &  270$\le$T$\le$5000 &  3.35476$\times10^{-6 }$ & -0.584062 & 4525.43 & 0.0073  \\
	&	2    & 400$\le$T$\le$1000 & 9.13678$\times10^{-7 }$ & -0.553899 & 6100.88 & 0.0021   \\
	&	      & 1000$<$T$\le$5000 & 3.04596$\times10^{-7 }$ & -0.420925 & 5907.02 & 0.0028   \\
	&	3    & 600$\le$T$\le$1500 & 9.67070$\times10^{-7 }$ & -0.509170 & 3228.55 & 0.0062   \\
	&	     & 1500$<$T$\le$5000 & 4.98432$\times10^{-6 }$ & -0.699380 & 9626.89 & 0.0040   \\
	&	4    & 850$\le$T$\le$2000 & 9.30628$\times10^{-8 }$ & -0.292379 & 11989.3 & 0.0054   \\
	&	     & 2000$<$T$\le$5000 & 8.81469$\times10^{-7 }$ & -0.546342 & 12651.0 & 0.0033   \\
	&	5    & 1000$\le$T$\le$5000 & 8.92493$\times10^{-7 }$ & -0.596125 & 14871.8 & 0.0066  \\
	&	6    & 1200$\le$T$\le$2500 & 1.44702$\times10^{-6 }$ & -0.667566 & 17980.5 & 0.0022   \\
	&	     & 2500$<$T$\le$5000 & 5.49126$\times10^{-6 }$ & -0.815886 & 18422.2 & 0.0009   \\
	&	7    & 1500$\le$T$\le$5000 & 1.65552$\times10^{-6 }$ & -0.696318 & 21345.0 & 0.0190  \\
	&	8    & 1800$\le$T$\le$5000 & 9.86709$\times10^{-8 }$ & -0.470871 & 23504.3 & 0.0071  \\
	&	9    & 2000$\le$T$\le$5000 & 1.34002$\times10^{-6 }$ & -0.780327 & 26309.7 & 0.0025  \\
		\hline
 1 	&	{\bf 0}    & 100$\le$T$\le$5000 & 1.45341$\times10^{-6 }$ & -0.478820 & 19.3876 & 0.0195   \\
	&	2    & 190$\le$T$\le$900 & 3.82842$\times10^{-6 }$ & -0.706740 & 3146.96 & 0.0154    \\
	&	     & 900$<$T$\le$5000 & 1.45126$\times10^{-8 }$ & -0.015912 & 2326.31 & 0.0076    \\
	&	3    & 420$\le$T$\le$900 & 1.79064$\times10^{-7 }$ & -0.420497 & 6137.15 & 0.0011   \\
	&	      & 900$<$T$\le$5000 & 1.14329$\times10^{-8 }$ & -0.080877 & 5736.89 & 0.0040   \\
	&	4    & 600$\le$T$\le$1500 & 5.53686$\times10^{-8 }$ & -0.277018 & 8726.55 & 0.0040     \\
	&	      & 1500$<$T$\le$5000 & 4.12085$\times10^{-7 }$ & -0.506359 & 9262.99 & 0.0065     \\
	&	5    & 850$\le$T$\le$5000 & 1.38279$\times10^{-7 }$ & -0.426255 & 12008.4 & 0.0031  \\
	&	6    & 1100$\le$T$\le$2500 & 9.87396$\times10^{-7 }$ & -0.707100 & 14813.1 & 0.0035  \\
	&	      & 2500$<$T$\le$5000 & 3.02402$\times10^{-8 }$ & -0.318434 & 13684.3 & 0.0017  \\
	&	7    & 1300$\le$T$\le$5000 & 3.25083$\times10^{-8 }$ & -0.336661 & 16732.0 & 0.0157  \\
	&	8    & 1500$\le$T$\le$5000 & 1.39545$\times10^{-7 }$ & -0.467278 & 20348.3 & 0.0062  \\
	&	9    & 1800$\le$T$\le$5000 & 3.50537$\times10^{-8 }$ & -0.362041 & 22692.0 & 0.0027  \\
		\hline
2	&	{\bf 1}    & 100$\le$T$\le$1500 & 1.04603$\times10^{-7 }$ & -0.287973 & -9.07849 & 0.0073  \\
 	&	             & 1500$<$T$\le$5000 & 4.35037$\times10^{-9 }$ & 0.116715 & -377.074 & 0.0027  \\
	&	{\bf 0}    & 100$\le$T$\le$5000 & 6.63440$\times10^{-7 }$ & -0.498417 & 38.4238 & 0.0193   \\
	&	3    & 200$\le$T$\le$700 & 1.79199$\times10^{-7 }$ & -0.415799 & 2977.42 & 0.0005  \\
	&	      & 700$<$T$\le$5000 & 7.47837$\times10^{-8 }$ & -0.306566 & 2843.67 & 0.0089  \\
	&	4    & 400$\le$T$\le$1300 & 4.27598$\times10^{-6 }$ & -0.800392 & 6144.90 & 0.0036  \\
	&	      & 1300$<$T$\le$5000 & 4.37539$\times10^{-8 }$ & -0.268995 & 5056.36 & 0.0148  \\
	&	5    & 600$\le$T$\le$1500 & 7.02948$\times10^{-8 }$ & -0.367691 & 8687.59 & 0.0076  \\
	&	      & 1500$<$T$\le$5000 & 8.34212$\times10^{-9 }$ & -0.117668 & 8217.86 & 0.0022  \\
	&	6    & 850$\le$T$\le$5000 & 9.67864$\times10^{-8 }$ & -0.401678 & 11429.0 & 0.0164  \\
	&	7    & 1000$\le$T$\le$2500 & 9.38561$\times10^{-7 }$ & -0.733592 & 14527.8 & 0.0013  \\
	&	     & 2500$<$T$\le$5000 & 1.75259$\times10^{-7 }$ & -0.547538 & 13960.8 & 0.0012  \\
	&	8    & 1300$\le$T$\le$5000 & 1.13024$\times10^{-6 }$ & -0.737221 & 17518.6 & 0.0117  \\
	&	9    & 1500$\le$T$\le$5000 & 1.46292$\times10^{-7 }$ & -0.513870 & 19732.3 & 0.0137  \\
		\hline
3	&	{\bf 2}    & 100$\le$T$\le$5000 & 3.60161$\times10^{-7 }$ & -0.490909 & 23.3038 & 0.0108  \\
	&	{\bf 1}    & 100$\le$T$<$700 & 2.52673$\times10^{-9 }$ & 0.215750 & -58.7472 & 0.0174  \\
	&	             & 700$\le$T$<$5000 & 7.86265$\times10^{-8 }$ & -0.292543 & 41.8077 & 0.0060  \\
%	&	{\bf 0}    & 100$\le$T$\le$700 & 3.69265$\times10^{-7 }$ & -0.399263 & -29.4462 & 0.0087  \\
		\hline
	\end{tabular}
%	\normalsize
%\vspace{-0.5cm}
%\end{table}
%\begin{table*}[t]
%	\renewcommand{\arraystretch}{1.00}
%	\centering
%	\caption{\label{tab:4}List of the parameters used in Eq.~(\ref{eqn:N2_VE_VdE_Interpolation}) for the VE and VdE rate coefficients of N$_2^+$  ($v_i^+=0-5$). The lines having bold $v_f^+$ values belong to VdE.}
%	\footnotesize
%	\scriptsize
%	\vspace{-0.5cm}
	\begin{tabular}{ccccccc}
		\hline
	$v_i^+$ &	$v_f^+$ & Temperature range & $A$             & $\alpha$         & $B$ & RMS \\
	 &	 &   (K) & $(cm^3s^{-1}K^{-\alpha})$   &      & $(K)$ & \\
		\hline
%3	&	{\bf 2}    & 100$\le$T$\le$5000 & 3.60161$\times10^{-7 }$ & -0.490909 & 23.3038 & 0.0108  \\
%	&	{\bf 1}    & 100$\le$T$<$700 & 2.52673$\times10^{-9 }$ & 0.215750 & -58.7472 & 0.0174  \\
%	&	             & 700$\le$T$<$5000 & 7.86265$\times10^{-8 }$ & -0.292543 & 41.8077 & 0.0060  \\
3	&	{\bf 0}    & 100$\le$T$\le$700 & 3.69265$\times10^{-7 }$ & -0.399263 & -29.4462 & 0.0087  \\
	&	             & 700$<$T$\le$5000 & 1.88955$\times10^{-6 }$ & -0.594995 & 205.425 & 0.0045  \\
	&	4    & 100$\le$T$\le$560 & 3.67969$\times10^{-7 }$ & -0.546948 & 2807.96 & 0.0290  \\
	&	      & 560$<$T$\le$5000 & 1.29198$\times10^{-7 }$ & -0.410534 & 2682.71 & 0.0112  \\
	&	5    & 440$\le$T$\le$1300 & 4.82997$\times10^{-7 }$ & -0.335457 & 5887.42 & 0.0106  \\
	&	      & 1300$<$T$\le$5000 & 8.49983$\times10^{-9 }$ & -0.137448 & 5427.34 & 0.0084  \\
	&	6    & 650$\le$T$\le$1700 & 1.54145$\times10^{-7 }$ & -0.532191 & 9039.15 & 0.0032  \\
	&	      & 1300$<$T$\le$5000 & 9.06525$\times10^{-10 }$ & 0.052019 & 7617.98 & 0.0111  \\
	&	7    & 850$\le$T$\le$2000 & 2.49601$\times10^{-8 }$ & -0.300658 & 11333.8 & 0.0091  \\
	&	      & 2000$<$T$\le$5000 & 4.15424$\times10^{-9 }$ & -0.093853 & 10888.7 & 0.0003  \\
	&	8    & 1100$\le$T$\le$5000 & 2.27558$\times10^{-9 }$ & -0.079760 & 13723.8 & 0.0189  \\
	&	9    & 1300$\le$T$\le$5000 & 3.06856$\times10^{-7 }$ & -0.645208 & 17278.8 & 0.0034  \\
		\hline
4	& 	{\bf 3}    & 100$\le$T$\le$400 & 6.37863$\times10^{-8 }$ & -0.145968 & -28.4720 & 0.0082  \\
 	& 	             & 400$<$T$\le$5000 & 6.36181$\times10^{-6 }$ & -0.825697 & 192.216 & 0.0072  \\
	&	{\bf 2}    & 100$\le$T$\le$900 & 6.55402$\times10^{-8 }$ & -0.262797 & 23.6274 & 0.0227  \\
	&	             & 900$<$T$\le$5000 & 1.05821$\times10^{-7 }$ & -0.378427 & -214.371 & 0.0140  \\
	&	{\bf 1}    & 100$\le$T$\le$440 & 2.32972$\times10^{-8 }$ & -0.116666 & -8.97262 & 0.0057  \\
	&	             & 440$\le$T$\le$5000 & 2.27097$\times10^{-7 }$ & -0.444141 & 124.096 & 0.0044  \\
	&	{\bf 0}    & 100$\le$T$\le$400 & 2.12911$\times10^{-9 }$ & 0.389295 & -22.4417 & 0.0183  \\
	&	             & 400$<$T$\le$5000 & 2.10016$\times10^{-6 }$ & -0.645580 & 264.431 & 0.0066  \\
	&	5    & 190$\le$T$\le$900 & 7.69858$\times10^{-7 }$ & -0.525187 & 2964.92 & 0.0217  \\
	&	      & 900$<$T$\le$5000 & 2.43298$\times10^{-6 }$ & -0.666480 & 3146.71 & 0.0028  \\
	&	6    & 400$\le$T$\le$1200 & 4.49741$\times10^{-6 }$ & -0.896475 & 5955.41 & 0.0017  \\
	&	      & 1200$<$T$\le$5000 & 8.84487$\times10^{-7 }$ & -0.708372 & 5563.92 & 0.0076  \\
	&	7    & 580$\le$T$\le$1500 & 1.17991$\times10^{-5 }$ & -0.974002 & 8780.06 & 0.0036  \\
	&	      & 1500$<$T$\le$5000 & 7.34677$\times10^{-7 }$ & -0.655786 & 8053.50 & 0.0084  \\
	&	8    & 800$\le$T$\le$2000 & 4.10791$\times10^{-6 }$ & -0.886512 & 11434.1 & 0.0015  \\
	&	      & 2000$<$T$\le$5000 & 3.73224$\times10^{-7 }$ & -0.615368 & 10737.3 & 0.0026  \\
	&	9    & 1000$\le$T$\le$2500 & 1.13754$\times10^{-7 }$ & -0.519202 & 13677.4 & 0.0126  \\
	&	      & 2500$<$T$\le$5000 & 1.25100$\times10^{-8 }$ & -0.269595 & 13039.9 & 0.0002  \\
		\hline
5	&	{\bf 4}    & 100$\le$T$\le$5000 & 4.50970$\times10^{-7 }$ & -0.474388 & -11.0566 & 0.0125  \\
	&	{\bf 3}    & 100$\le$T$\le$500 & 5.83098$\times10^{-8 }$ & -0.304037 & -19.8859 & 0.0073  \\
	&	             & 500$<$T$\le$5000 & 1.31781$\times10^{-7 }$ & -0.447966 & -55.6242 & 0.0132  \\
	&	{\bf 2}    & 100$\le$T$\le$5000 & 6.21686$\times10^{-8 }$ & -0.340696 & -27.4858 & 0.0101  \\
	&	{\bf 1}    & 100$\le$T$\le$900 & 8.49509$\times10^{-8 }$ & -0.310555 & -17.4355 & 0.0022 \\
	&	             & 900$<$T$\le$5000 & 1.08934$\times10^{-6 }$ & -0.653000 & 198.309 & 0.0031 \\
	&	{\bf 0}    & 100$\le$T$\le$5000 & 3.73256$\times10^{-7 }$ & -0.501538 & 2.80363 & 0.0122 \\
	&	6    & 190$\le$T$\le$5000 & 1.47602$\times10^{-7 }$ & -0.367489 & 2789.39 & 0.0150 \\
	&	7    & 400$\le$T$\le$1200 & 3.15877$\times10^{-8 }$ & -0.292229 & 5543.78 & 0.0036 \\
	&	      & 1200$<$T$\le$5000 & 1.10525$\times10^{-6 }$ & -0.712115 & 6289.63 & 0.0090 \\
	&	8    & 600$\le$T$\le$5000 & 4.10724$\times10^{-7 }$ & -0.585006 & 8593.14 & 0.0153 \\
	&	9    & 850$\le$T$\le$2000 & 7.61666$\times10^{-7 }$ & -0.744946 & 11307.6 & 0.0038 \\
	&	      & 2000$<$T$\le$5000 & 7.87895$\times10^{-8 }$ & -0.489941 & 10618.0 & 0.0036 \\
		\hline
	\end{tabular}
	\normalsize
\end{table*}

%The rate coefficients of the vibrational transitions (VE and VdE) of N$_2^+$ have been interpolated using the similar  formula:
%
%\begin{equation}\label{eqn:N2_VE_VdE_Interpolation}
%k_{v_i^+\to v_f^+}^{VE,VdE}(T_e) = A_{v_i^+\to v_f^+} \, T_e^{\alpha_{v_i^+\to v_f^+}} \, \exp\left[-\frac{B_{v_i^+\to v_f^+}}{T_e}\right]
%\label{kve}
%\end{equation}
%
%\noindent over the electron temperature range 450~K~$\le T_e\le$~$~5~000~K$.  
%The parameters $ A_{v_i^+\to v_f^+}$, $\alpha_{v_i^+\to v_f^+}$ and $B_{v_i^+\to v_f^+}$ are listed in Table \ref{tab:N2_VE_VdE_Interpolation}. 
%
%A representative example of the relative differences between the fitted data and the original ones is presented in 
%fig.~\ref{fig:7}. A much more accurate fit has been produced  using generalized Arrhenius-type formulas ~\cite{Niyonzima2017}, and the corresponding fitting parameters will be accessible in the Supplementary Materials. 

\section{Conclusions}

The present work extends considerably our previous study of the dissociative recombination of N$_2^+$ with electrons~\cite{Little2014}. Making use of the molecular data set calculated in Refs.~\cite{Little2014,Little-resonance_state, Little_bound_state} and of our step-wise MQDT method, we have performed calculations for the lowest $6$ vibrational levels of the target cation in collision with electrons having kinetic energy up 2.3 eV and, in the case of thermal equilibrium, electronic temperature up to 5000 K.
We have provided cross sections and rate coefficients for dissociative recombination, vibrational excitation and de-excitation of N$_2^+$ molecular cation, important for the detailed kinetic modelling of cold astrophysical, atmospheric and laboratory plasmas. The calculated cross sections and rate coefficients are available on request.

\begin{acknowledgments}
The authors acknowledge support from F\'ed\'eration de Recherche Fusion par Confinement Magn\'etique (CNRS and CEA), La R\'egion Normandie, FEDER and LabEx EMC3 via the projects PTOLEMEE, Bioengine, EMoPlaF, COMUE Normandie Universit\'e, the Institute for Energy, Propulsion and Environment (FR-IEPE), the European Union via COST (European Cooperation in Science and Technology) action MD-GAS (CA18212), and ERASMUS-plus conventions between Universit\'e Le Havre Normandie and University College London.
This work has received funding from the Euratom research and training programme 2014-2018 and 2019-2020 under grant agreement No. 633053. The views and opinions expressed herein do not necessarily reflect those of the European Commission.
We are indebted to Agence Nationale de la Recherche (ANR) via the project MONA, Centre National de la Recherche Scientifique via the GdR TheMS and the DYMCOM project, and the Institute Pascal, University Paris-Saclay for the warm hospitality during the DYMCOM workshop. 
This work was supported by the Programme National 'Physique et
Chimie du Milieu Interstellaire' (PCMI) of CNRS/INSU with INC/INP co-funded
by CEA and CNES.
JZsM thanks the financial support of the National Research, Development and Innovation Fund of Hungary, under the K 18 funding scheme with project no. K128621.
\end{acknowledgments}

\section*{Data Availability}
The data that support the findings of this study are available from the corresponding author upon reasonable request.

\nocite{*}
\section*{References}
\bibliography{Schneider-N2-4arxiv.bib}% Produces the bibliography via BibTeX.

\end{document}